\documentclass{article}
\usepackage{spconf,amsmath,graphicx}
\usepackage{amsfonts}
\usepackage{multirow}
\usepackage{subcaption}

\newcommand{\oOr}[1]{\mathbf{O}^{(#1)}}

\newcommand{\wWr}[1]{W^{(#1)}}

\newcommand{\pr}{\mathbb{P}}
\newcommand{\acc}{\mathbb{A}}

\newcommand{\blank}{$\phi$}

\newcommand{\NumG}{\mathcal{G}_\text{Num}}
\newcommand{\DenG}{\mathcal{G}_\text{Den}}

\newcommand{\Flfbmmi}{\mathcal{F}_\text{LF-bMMI}}

\usepackage{tablefootnote}
\usepackage{placeins}
\usepackage{float}
\usepackage{tikz}
\usepackage{subcaption}
\usetikzlibrary{shapes,arrows,automata,positioning}
\usepackage{environ}

\NewEnviron{FitToWidth}[1][\textwidth]{%
  \makebox[#1]{\resizebox{#1}{!}{\BODY}}%
}

\title{On lattice-free boosted MMI training of HMM and CTC-based full-context ASR models}

\name{
\begin{tabular}{c}
Xiaohui Zhang$^{\star}$, Vimal Manohar$^{\star}$, David Zhang, Frank Zhang, Yangyang Shi,  \\ Nayan Singhal, Julian Chan,
Fuchun Peng, Yatharth Saraf, Mike Seltzer \thanks{The authors would like to thank Dan Povey, Yiming Wang, Yiwen Shao, Michael Picheny, and Zhijian Ou for insightful discussions and suggestions.}
\end{tabular}
\thanks{$\star$ Equal contribution.}  }
\address{Facebook AI, USA}
\begin{document}
\ninept
\maketitle
\begin{abstract}
Hybrid automatic speech recognition (ASR) models are typically sequentially trained with CTC or LF-MMI criteria. However, they have vastly different legacies and are usually implemented in different frameworks. In this paper, by decoupling the concepts of modeling units and label topologies and building proper numerator/denominator graphs accordingly, we establish a generalized framework for hybrid acoustic modeling (AM). In this framework, we show that LF-MMI is a
powerful training criterion applicable to both limited-context and full-context models, for wordpiece/mono-char/bi-char/chenone units, with both HMM/CTC topologies. From this framework, we propose three novel training schemes: chenone(ch)/wordpiece(wp)-CTC-bMMI, and wordpiece(wp)-HMM-bMMI with different advantages in training performance, decoding efficiency and decoding time-stamp accuracy. The advantages of different training schemes are evaluated comprehensively on Librispeech, and wp-CTC-bMMI and ch-CTC-bMMI are evaluated on two real world ASR tasks to show their effectiveness. Besides, we also show bi-char(bc) HMM-MMI models can serve as better alignment models than traditional non-neural GMM-HMMs. 
\end{abstract}
\vspace{-0.5em}
\section{Introduction}
\label{sec:intro}
\vspace{-0.5em}
State-of-the-art Automatic Speech Recognition (ASR) systems use Deep Neural Networks (DNN) of various architectures for acoustic modeling (AM). Early success using DNNs for ASR came from hybrid DNN-hidden markov models (DNN-HMM) \cite{hmm_dnn}. These were typically trained using frame-level cross entropy (CE) criterion to predict senones \cite{hmm_dnn} obtained from a previous Gaussian Mixture Model(GMM)-HMM system. Sequence-level training criteria like Maximum Mutual Information (MMI) \cite{bahl_mmie} have been shown to improve the performance of these frame-level trained DNN-HMM-based ASR systems \cite{Vesely_13, saon2012discriminative}. Since then, various approaches have been shown to be able to train neural network purely through sequence training without initially pre-training using a frame-level criterion -- lattice-free MMI (LF-MMI) \cite{lfmmi, e2e_lfmmi}, connectionist temporal classification (CTC) \cite{ctc}, recurrent neural network transducer (RNN-T) \cite{graves2012rnnt}, attention-based sequence-to-sequence (seq2seq) models \cite{bahdanau2016attention, las}.

RNN-T and seq2seq models consist of an acoustic encoder that is jointly trained with a neural decoder, which can be considered to be a neural language model (LM). These models can be used to decode audio without using an external LM, and thus can be termed as ``end-to-end". As opposed to this, CTC-based models and hybrid DNN-HMM are ``encoder-only" models in the sense that they do not have an explicit jointly trained neural decoder. 
Having a single ``end-to-end" model might be simpler, but in general these models are known to be data-hungry \cite{chiu2018seq2seq, xiaohui2021benchmarking} and require thousands of hours of data to achieve competitive performance. RNN-T models are also known to benefit from pre-training encoders or alignments from CTC \cite{graves2013rnn} or hybrid DNN-HMM \cite{chunxi2021aux} models for accuracy or efficiency improvements \cite{graves2013rnn,chunxi2021aux,ar_rnnt, Zeyer2020transducer}. On the other hand, hybrid models use an external LM for decoding and are often explicitly trained to work with an LM \cite{lfmmi,semisup_lfmmi,ctc_crf}. They are appealing for their modularity which allows to easily replace or extend the lexicon or LM for different applications, while this is still a challenge for end-to-end systems \cite{duc2021rnnt_context}. Hybrid models also explicitly model silence which makes them ideal candidates for pre-processing and segmenting audio as well as for applications that require highly accurate decoding token time-stamps.

While hybrid DNN-HMM and CTC models are very similar, they have vastly different legacies and are usually implemented in very different frameworks. 
For e.g., though LF-MMI was proposed in a DNN-HMM framework with senone/chenone\cite{le2019senones} modeling units, this combination of topology and modeling units is not mandatory. On the other hand, CTC models conventionally refers to a model whose modeling units follow the CTC topology and trained with the Maximum-Likelihood (ML) criteria, which is just the numerator part of the MMI criteria \cite{e2e_lfmmi}. However, CTC models can also be trained discriminatively with sMBR \cite{ctc_smbr}, or MMI~\footnote{The CTC-CRF criterion in \cite{ctc_crf} is equivalent to LF-MMI as in \cite{e2e_lfmmi} as both used uniform transition scores constant over the linear chain.} criteria. Intrinsically HMM and CTC are just different label topologies (Sec. \ref{sec:topo}). By decoupling the concepts of modeling units (character/wordpiece/chenone etc.) and label topologies, we introduce a single generalized framework for training hybrid models. This major contribution of our paper allows systematic comparisons (Sec. \ref{libri_setup}) of different modeling units and label topologies to gain deep understandings of their properties, and makes it easier to develop training schemes with novel combinations of them.


 
From this framework, together with the boost factor \cite{povey2008bmmi,chen2018sequence} for LF-MMI, we propose three new training schemes: 1,2) wp-CTC-bMMI and ch-CTC-bMMI (CTC-bMMI with chenone/wordpiece units), with overall better WERs than HMM-bMMI, whose effectiveness is also confirmed by two real-world server-side/on-device ASR applications. 3) wp-HMM-bMMI, which enables both large-stride (8) inference and accurate token time-stamps, thanks to silence modeling. On the HMM side, we also show HMM-MMI models with bi-character units (bc-HMM-MMI) can serve as a better flat-start trained alignment model than Gaussian Mixture Models (GMM), especially on noisy data.

\vspace{-0.5em}
\section{LF-bMMI training}
\vspace{-0.5em}
LF-MMI \cite{lfmmi} criterion was extended to include boosting \cite{povey2008bmmi} in \cite{chen2018sequence, weng2019lfbmmi}. Here, we present it again in the generalized hybrid model framework for different modeling units and label topologies.

The MMI criterion \cite{bahl_mmie} for training acoustic models can be viewed as maximizing the conditional likelihood of the reference $\wWr{r}$ given the acoustic observation sequence $\oOr{r}$. This maximizes the joint likelihood of the reference and acoustic observation sequence, i.e. numerator likelihood, and minimizes the marginal likelihood $\oOr{r}$, i.e. denominator likelihood. As in \cite{lfmmi}, the denominator is approximated by marginalizing over all state sequences in a denominator graph $\DenG$ (hence ``lattice-free") constructed using an n-gram token LM, which in our case can be phone/character/wordpiece LM. The numerator likelihood is computed by marginalizing over all sequences in a numerator graph $\NumG(\wWr{r})$ that is similar but constrained to the reference word sequence. In this paper, we assume MMI/bMMI training is always lattice-free, hence omitting ``LF" most of the time. 

The boosted MMI \cite{povey2008bmmi,Vesely_13} criterion was introduced to improve training performance by encouraging the criterion to give higher likelihoods to more ``accurate" paths. This is achieved by boosting the likelihoods of paths in the denominator graph proportional to the number of errors it contains. The LF-bMMI criterion can be written as:

\begin{equation}
    \begin{aligned}
  \Flfbmmi =& \sum_{r} \log \frac{\sum_{\pi\in\NumG(\wWr{r})}
  \pr{\left(\oOr{r}\mid \pi\right)}^\kappa \pr(\pi)} {\sum_{\pi' \in \DenG}
  \pr{\left(\oOr{r}\mid \pi'\right)}^\kappa \pr(\pi') e^{-b\acc(\wWr{r},\pi')}}, \label{eq:Flfbmmi}
    \end{aligned}
\end{equation}
where $\kappa$ is acoustic weight and $\acc(\wWr{r},\pi')$ is the accuracy function for the path $\pi'$ measured against the reference $\wWr{r}$. The accuracy function can be defined in several ways such as using phone edit distance to the reference \cite{povey2008bmmi}. But implementation-wise, in the lattice-free training framework, it is easiest to define this as a sum of per-frame accuracy values. Therefore, as in \cite{weng2019lfbmmi}, we use numerator posterior derived from the numerator graph as a proxy for the per-frame state-level accuracy values. Besides, the intuition of boosted MMI can also be interpreted by Max-Margin learning \cite{baskar2019promising} \cite{smithsoftmax}.

\vspace{-0.5em}
\subsection{Full-sequence training}
\vspace{-0.5em}
The LF-(b)MMI criterion was originally designed at the sequence-level. For efficiency on GPUs, the original Kaldi implementation \cite{lfmmi} applies it on equally-sized chunks of around 1.5s each. However, in our application we need to apply LF-bMMI criterion to full-context models like BLSTMs and Transformers, and sequence lengths of up to 2 minutes. We leveraged PyChain's LF-MMI implementation for sequence-training with variable length sequences, and added boosting \cite{povey2008bmmi} for training with boosted MMI.
\vspace{-0.5em}
\section{Label topologies and modeling units}
\begin{figure}
\begin{subfigure}[h] {0.42\textwidth}
    \centering {
    \begin{FitToWidth}[0.88\columnwidth]
    \includegraphics[]{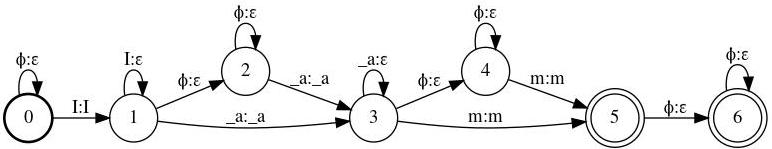}
    \end{FitToWidth}
  } 
  \caption{CTC topology\label{fig:ctc_num_fst}}
  \end{subfigure}
  
  \begin{subfigure}[h] {0.47\textwidth}
    \centering {
   \begin{FitToWidth}[1\columnwidth]
    \includegraphics[]{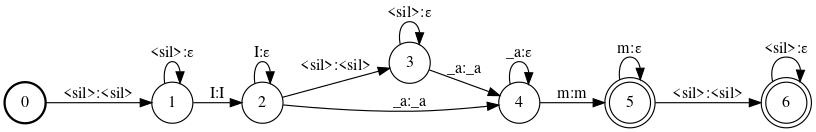}
    \end{FitToWidth}
  }
    \caption{\label{fig:1_state_num_fst}1-state HMM topology}
  \end{subfigure}
  
    \caption{Numerator FSTs (mapping output units to modeling units) of `I', `$\_$a', `m' in CTC and 1-state HMM topology; $\phi$ means blank.}
  
    \label{fig:topo}
    
\end{figure}

\vspace{-0.5em}
\label{sec:topo}
In this section, we describe the label topologies and modeling units used in our models. A label topology defines the mapping between a label sequence and neural network output units (i.e. modeling units). For DNN-HMM systems, in this paper, we consider only the 1-state and 2-state-with-skip (which we call as {\em chain}) HMM topologies \cite{e2e_lfmmi}. For CTC systems, a CTC topology \cite{ctc, ctc_crf} is used which adds a special blank (\blank) output unit. The CTC topology defines a mapping $\mathcal{B}^{-1}$ that maps a label sequence $\boldsymbol{l} = l_1,\dots l_L$ to all output unit sequences $\pi$ such $l$ is obtained by de-duplicating $\pi$ and removing blank symbols. An intuitive understanding of the difference between CTC and 1-state HMM topology\footnote{The {\em chain} topology can be obtained from Fig.\ref{fig:1_state_num_fst} by replacing input tokens on all self-loops by a `2nd version' of each token (e.g. `I'$\rightarrow$`I$_2$').} can be obtained by looking at the examples in Fig.\ref{fig:topo}. We see that the CTC topology allows blank ($\phi$) units {\em between any tokens}, e.g. \texttt{\_a} and \texttt{m}. The silence label (\texttt{<sil>}) which we see in the 1-state HMM topology is different from blank in that it is a real label similar to any other wordpiece. In our systems, we make the modeling choice to optionally allow it {\em between words} in order to model the real acoustics of silence \cite{chen2015pronunciation}.
We also point out that explicit silence modeling can help achieve more accurate token time-stamps during decoding, which is an advantage of HMM-based models, especially when time-constraints are used in training targets. Notably, we can use both blank and silence in the same model, which is the case for chenone-CTC models as pointed out in Table \ref{tab:combination}.  On the other hand, we hypothesize that CTC-based models can achieve better training performance as it benefits more from \textit{SpecAugment} due to the blank tokens, verified in our experiments. The blank tokens, which signify ``no output" are ideal to represent the perturbations due to feature masking, while HMM-based models are forced to model masked features using non-silence units (except between words where silence can be predicted). However the cost is less accurate decoding time-stamps due to the peaky behavior \cite{zeyer2021does} caused by dominance of blank tokens at output during decoding.


We consider the following 4 types of labels:

\begin{itemize}
    \vspace{-0.5em}
    \item Mono-character (mono-char): 
    This is the simplest case where the labels are characters which are context-independent. 
    \vspace{-0.5em}
    \item Bi-character (bi-char):
    In this case, the characters are modeled separately for each left-context. We do a basic text-based clustering based on the raw counts of the character n-grams seen in the training transcripts, to let infrequent bi-characters share a modeling unit within each cluster. 
        \vspace{-0.5em}

    \item Tri-character (tri-char):
    In this case, the characters are modeled separately for each left and right context. We use standard decision-tree based clustering of states \cite{young_tied_states} and share modeling units across states within each cluster. We refer to tri-char based modeling units as chenones \cite{le2019senones}.
    
        \vspace{-0.5em}

    \item Wordpiece (wp):
    In this case, wordpieces are constructed using Sentencepiece \cite{spm} modeling from training transcripts.
\end{itemize}
\vspace{-0.5em}

\section{Numerator and Denominator preparation}
\vspace{-0.5em}
\subsection{HMM topology}
\vspace{-0.5em}
\textbf{\textit{Chenone units:}} We use the approach for denominator graph preparation from \cite{lfmmi}, except for replacing phoneme with character i.e. by composing an \{3,4\}-gram character LM with tri-character context-dependency transducer and HMM transducer. The n-gram LM was estimated using the alignments from a previous flatstart trained hybrid LFMMI bi-char system \cite{e2e_lfmmi}. Numerator graph preparation also follows the same approach from \cite{lfmmi} and we apply time-constraints using alignments from the same flatstart-trained hybrid system.

\noindent \textbf{\textit {Bi/mono-char units:}} The denominator graph preparation follows the similar approach as described for chenone units in the previous section but using a bi-character context-dependency for the bi-char systems and no context-dependency for the mono-char systems. Also the character LM has to be estimated from transcripts rather than alignments, with randomly inserted silence phones \cite{e2e_lfmmi}. Numerator graph in this case is a full HMM with self-loops following \cite{e2e_lfmmi}.

\noindent \textbf{\textit {Wordpiece units:}} The denominator and numerator graph preparation mimic the approach for mono-char units as described in the previous section. The word sequences are converted to wordpieces using a ``wordpiece lexicon" constructed using mappings from a Sentencepiece model \cite{kudo2018sentencepiece} trained on the text. Since the number of wordpieces is usually much larger than the number of characters, to decrease the denominator graph size, we use a \{2,3\}-gram LM on wordpieces for the denominator. An example for numerator FST for wordpiece units with HMM topology is shown in Figure \ref{fig:1_state_num_fst}.
\vspace{-0.5em}
\subsection{CTC topology}
\vspace{-0.5em}
\textbf{\textit{Chenone units:}} For chenone units with CTC topology, we first obtain the chenone sequence from a previous flatstart trained hybrid LFMMI bi-char system as in the case of HMM topology described in the chenone-HMM case, and remove repetitions to obtain a label sequence with chenones as the labels. We treat chenones similar to regular characters and compose the sequence with the CTC topology transducer \cite{ctc_crf}.
Note that unlike the chenone HMM case, there's no time constraints on numerator FSTs here. For the denominator graph, we first obtain the denominator graph for a 1-state HMM topology as in HMM case, and then convert it to a CTC compatible topology by splitting each state into two states and adding two arcs for consuming blank tokens, in the same way as done in  \cite{zhang2020wp_hybrid} for constructing decoding graph for chenone-based CTC models.

\noindent  \textbf{\textit {Wordpiece units:}} For wordpiece units with CTC topology, the numerator graph (e.g. \ref{fig:1_state_num_fst}) is created in the same way as the chenone case. The wordpiece sequence is generated on-the-fly by tokenizing the reference sequence into wordpieces 
using a SentencePiece model \cite{kudo2018sentencepiece}. The denominator graph is created by composing a n-gram wordpiece LM with the CTC topology transducer. The n-gram wordpiece LM is estimated from training transcripts tokenized into wordpieces using a Sentencepiece model. 

We summarized the main properties of the combinations of HMM/CTC topology with different modeling units which we'll study in Table \ref{combination}. Among them, wp-HMM, wp-CTC and ch-CTC are novel schemes in terms of MMI training. 

\begin{table*}[t]
\renewcommand\arraystretch{1.2}
    \centering
        \caption{Properties of combinations of different modeling units and label topologies (`mc' = `mono-char', `bc' = `bi-char', `ch'=`chenone')}
        \vspace{-0.2cm}
    \begin{tabular}{ || c || c | c | c | c | c ||}
    \hline
    \hline
Model & wp-HMM & mc/bc-HMM & ch-HMM & ch-CTC &  wp-CTC \\
\hline
\hline
Label topology & \multicolumn{3}{c|}{HMM} & \multicolumn{2}{c||}{CTC} \\
\hline
Acoustic-based clustering   & \multicolumn{2}{c|}{N} &\multicolumn{2}{c|}{Y} & N \\
\hline 
Time-constrained Num. FST & \multicolumn{2}{c|}{N}  & Y & \multicolumn{2}{c||}{N} \\
\hline 
Explicit silence modeling &  \multicolumn{4}{c|}{Y} & N \\ 
\hline
Training criterion & \multicolumn{2}{c|}{ML / MMI} & CE / MMI &  \multicolumn{2}{c||}{ML / MMI} \\
\hline
\hline
\end{tabular}

\label{combination}
\label{tab:combination}
\end{table*}
\vspace{-0.5em}
\section{Pre-training with CE/ML models}
\vspace{-0.5em}
To improve LF-bMMI training performance, we can pre-train the model with either frame-level CE criterion or sequence level ML criterion \cite{e2e_lfmmi}\footnote{Strictly speaking CE is frame-level ML. We make CE comparable to ML since we always refer ML to ``sequence-level ML" in our paper for simplicity.}. ch-HMM models (i.e. HMM topology with chenone units) are the only one for which we use frame-level alignments. For these, we use CE pre-training with the labels obtained from frame-level alignments. For other models, we use sequence-level pre-training with ML criterion. Note that in the case of CTC topology, this is equivalent to the CTC training criterion. In all these cases, the neural network outputs are locally normalized by a softmax layer.

When fine-tuning a neural network pre-trained with CE or ML criterion, we empirically found removing softmax and using the logits directly helped performance. 
However, we subtract the log of the model priors from the logits just as we would when using the model for decoding \cite{hmm_dnn}. We estimate the model priors \cite{manohar2015semi} on a small subset of training data as opposed to the conventional approach of obtaining it from frame-level alignments \cite{hmm_dnn}. This approach is more general as it allows to estimate model priors even for CTC-based systems with blank tokens and for wordpiece-based systems. We additionally apply an acoustic scale $\kappa$ on the neural network outputs before it is combined with the graph scores from the numerator or denominator graphs. In theory, the LF-bMMI objective is normalized at the sequence-level and hence it is capable of learning the linear offset corresponding to the log-priors as well as the acoustic scale. We indeed find that when the model is trained from scratch, we do not need to explicitly supply the log priors or an acoustic scale of 1.0 suffices. But when fine-tuning a pre-trained network, we found that we need to match the priors and acoustic scale to the optimal values during decoding. Using a mis-matched prior or acoustic scale leads to slower convergence.

\setlength{\tabcolsep}{0.14cm}
\begin{table*}[t]
\caption{\textit{dev-clean/other} ML/CE vs. MMI WER and the effect of ML/CE pre-training for MMI (\#ep means \# epochs to reach the best WER).} 
\begin{subtable}[h] {0.65\textwidth}
\centering
\small
\vspace{-0.2cm}
    \begin{tabular}{  c | c  c | c c | c c}
\multirow{2}{*}{Loss} & \multicolumn{2}{c}{wp-HMM} & \multicolumn{2}{|c}{wp-CTC} & \multicolumn{2}{|c}{ch-CTC}  \\
\cline{2-7}  
  &  WER &	 \#ep &  WER & \#ep &  WER & \#ep \\
\hline
ML & 7.2 / 17.3 & 	69 & 	4.6 / 11.5 & 	58 & 	4.1 / 10.7 & 	55 \\
MMI &  4.3 / 11.0 & 	60 & 	4.4 / 10.6  & 	121 & 	3.8 / 9.1 & 	153 \\
ML $\rightarrow$ MMI & 4.4 / 11.0 & 66 & 4.1 / 10.2 & 89 & 3.7 / 9.0 & 143 \\
\hline
\end{tabular}
\end{subtable}
\begin{subtable}[h] {0.1 \textwidth}
\centering
\small
\vspace{-0.2cm}
    \begin{tabular}{  c | c  c}
\multirow{2}{*}{Loss} & \multicolumn{2}{c}{ch-HMM}  \\
\cline{2-3}  
  &  WER & \#ep  \\
\hline
CE & 4.2 / 10.6 &	60 \\
MMI &  4.0 / 9.5 &	54 \\
CE $\rightarrow$ MMI & 3.8 / 9.1 & 48 \\
\hline
\end{tabular}
\end{subtable}
\label{ml}
\end{table*}
\vspace{-0.5em}
\section{Experiments}
\vspace{-0.5em}
\subsection{Comprehensive Analysis on Librispeech}
\vspace{-0.5em}
\label{libri_setup}
Here we perform a series of analysis of LF-bMMI training with different modeling units, label topologies and various configurations on Librispeech \cite{librispeech}. We use the standard (960h) training and (\textit{dev-clean}, \textit{dev-other} sets for training and evaluation respectively. We use the official 4-gram LM pruned to 3-gram with a threshold of $1e^{-9}$) built into HLG/HCLG graphs for decoding. For the AM, we use a 25M-parameters TDNN-BLSTM network with 2 BLSTM \cite{lstm} layers (640 hidden units) in each recurrence direction and 3 TDNN layers \cite{tdnn,vijay_tdnn} (640 hidden units) interleaved between input and first BLSTM layer, and between the 2 BLSTM layers. Unless specified, we use stride (i.e. input frame rate / output frame rate) 8 for wp-CTC/HMM and stride 4 for ch-CTC/HMM models, since previous studies \cite{zhang2020wp_hybrid} have shown wordpieces units can work reasonably well with stride 8, while chenone units cannot because of their short duration. Regarding modeling units, for mc-HMM, we use 29 characters. For bc-HMM, we use 870 bi-char units from text-based clustering. For ch-HMM/ch-CTC systems, we use a set of 1632 chenones corresponding to a tree built from alignments from a bc-HMM model. For wp-HMM/wp-CTC systems, we use a set of 511 wordpieces built from a Sentencepiece model, balancing performance between strides 4 and 8. Unless specified, we always conduct MMI training without pre-training, with $0$ as the boost factor, \texttt{LD} as the \textit{SpecAugment} policy, and 1-state topology for HMM-based systems.
\vspace{-0.5em}
\subsubsection{Basic results and the effect of ML/CE pre-training}
\vspace{-0.5em}
We first do comparison of the WERs of LF-MMI training for wp-HMM/CTC and ch-HMM/CTC with their corresponding non-discriminatively trained ML/CE baselines, and then investigate the effect of pre-training with ML/CE for LF-MMI training. Regarding the choice between ML/CE training, since ch-HMM is the only one with frame-level targets, it's natural to go with CE for ch-HMM, and ML for others. From results in Table \ref{ml}, comparing with ML baselines, we can see that wp/ch-CTC-MMI both have around $8-15\%$ relative improvements on \textit{dev-other} and $4-7\%$ relative improvements on \textit{dev-clean}, and pre-training MMI with ML helps provide a better initialization resulting in both faster convergence and better final WER. For wp-HMM, the ML WER is significantly worse and doesn't help for pre-training MMI, which is similar to the finding on mc-HMM in \cite{e2e_lfmmi}. For ch-HMM, MMI achieves $5-10\%$ improvement comparing with CE, and pre-training with CE further brings $4\%$ improvements. 
\vspace{-0.5em}
\subsubsection{The effect of boost}
\vspace{-0.5em}
Here we study the contribution of the boost factor for bMMI training. From Table \ref{boost} we can see that the boost improves WERs for all four systems. For wp-HMM, wp-CTC, ch-CTC, the relative WER gain is around $2-7\%$ on \textit{dev-clean} and $2-4\%$ on \textit{dev-other}. For ch-HMM, the gain is large: $10\%$ on \textit{dev-clean} and $6\%$ on \textit{dev-other}. We suspect the reason is that ch-HMM is the only system with time-constraints on the numerator FSTs, and thereby the frame posteriors are more accurate, which the boosting mechanism relies on.

\begin{table}[h]
    \centering
    \small
        \caption{\textit{dev-clean/other} bMMI WER with different boost values} 
        \vspace{-0.2cm}
    \begin{tabular}{ c || c | c | c | c }
boost & wp-HMM & wp-CTC &  ch-HMM & ch-CTC \\
\cline{1-5}  
\texttt{0}  & 4.3 / 11.0 & 4.4 / 10.6 & 4.0 / 9.5 &	3.8 / 9.1 \\
\texttt{0.3} & 4.2 / 11.0 & \textbf{4.2 / 10.4}	& 3.7 / 9.2	& 3.7 / 9.1 \\
\texttt{0.5} & \textbf{4.2 / 10.7} &	4.3 / 10.3 & \textbf{3.6 / 8.9}	&\textbf{ 3.6 / 8.7 }\\
\texttt{1.0} & 4.2 / 10.9 & 4.4 / 10.9 & 3.6 / 9.3 & 3.6 / 8.9 \\
\hline
\end{tabular}
\label{boost}
\end{table}
\vspace{-0.5em}
\subsubsection{The effect of \textit{SpecAugment}}
\vspace{-0.5em}
Here we study the effect of \textit{SpecAugment} for different systems. We study two \textit{SpecAugment} policies -- \texttt{LD}, \texttt{Large}. \texttt{LD} is same as in \cite{spec_augment} but with maximum time mask width of $p=0.2$. \texttt{Large} ($T=30, mT= 10$) is a more aggressive policy which was shown in \cite{zhang2020wp_hybrid} to help performance on Librispeech. From Table \ref{tab:specaug}, we see that without \textit{SpecAugment}, for both wordpiece and chenone units, HMM and CTC models have similar WERs. However, we see that CTC models benefit more from \textit{SpecAugment} compared to the corresponding HMM models, verifying our hypothesis on the advantage of CTC which better models feature masking with blank tokens.


\begin{table}[!h]
    \centering
    \small
        \caption{\textit{dev-clean/other} MMI WERs with different \textit{SpecAugment} policies} 
                \vspace{-0.2cm}
    \begin{tabular}{ p{1cm} || c | c | c | c }
Policy & wp-HMM & wp-CTC &  ch-HMM & ch-CTC \\
\cline{1-5}  
\texttt{None} & 4.8 / 13.0 & 4.8 / 13.1 & 4.5 / 11.6 & 4.5 / 11.7 \\
\texttt{LD} & 4.3 / 11.0 & 4.4 / 10.6 & 4.0 / 9.5 & 3.8 / 9.1 \\
\texttt{Large} & 4.4 / 10.8 & 4.3 / 10.3 & 3.9 / 9.2 & 3.8 / 8.9\\
\hline
\end{tabular}
\label{saug}
\label{tab:specaug}
\end{table}

\vspace{-0.5em}
\subsubsection{Comparing different modeling units}
\vspace{-0.5em}
Here we fix the label topology to be 1-state HMM, and compare the WER and RTF\footnote{When we measure RTF, we optimize the decoding beam so that the WER is $1\%$ worse than the optimal WER. Otherwise we always use a beam of 30.} performance of different modeling units, both wordpiece and character-based units.
For wordpiece, we train models with strides 8 and 4. For character-based units we couldn't get reasonable convergence performance with stride 8 and hence stick to stride 4. From Table \ref{units_perf}, we can see that the WER of bi-char is better than mono-char by a large gap ($13-15\%$ relative), while the relative improvement of tri-char on top of bi-char is smaller ($2-7\%$). This implies that even text-based simple clustering can provide quite useful context dependency information. Looking at wordpiece units, we can see that with stride 4, its performance is better than bi-char and close to tri-char, showing wordpieces can also be powerful modeling units without relying on decision tree building. Furthermore, at stride 8, we can see its performance is still $6-10\%$ better than mono-char at stride 4. Unfortunately, the RTFs we report here for wordpiece-based models are much worse than the mono-char case. This is due to increased number of modeling units (29 chars $\rightarrow$ 511 wordpieces), and hence more confusable paths during graph search. However, in real applications where we use much larger AMs so that AM inference dominates the computation, the RTF advantage of stride 8 wordpiece systems would amplified as verified in a previous study \cite{xiaohui2021benchmarking}, where a stride 8 wp-CTC model had better RTF than a stride 3 ch-HMM model using the same encoder.
\begin{table}[!h]
    \centering
    \small
        \caption{\textit{dev-clean/other} MMI WER, RTF and TSE of different units with the same (1-state) HMM topology}
    \begin{tabular}{  p{0.12\columnwidth} | c | c | c | c | c }
            \vspace{-0.2cm}
Unit & \multicolumn{2}{c|}{wordpiece} & mono-char & bi-char & chenone \\
\cline{1-6}
\hline
Stride & 8 & \multicolumn{4}{c}{4} \\
\hline
\hline
WER & 4.4 / 11.1 & 3.9 / 10.1 & 4.9 / 11.8 &	4.2 / 10.3 &	4.1 / 9.6 \\
\hline
RTF & 0.020 & 0.046 & 0.006 & 0.005 & 0.011 
  \\ \hline
TSE  & 86 & 66 & 74 & 47 & 28 \\
\hline
\end{tabular}
\label{units_perf}
\label{tab:units_perf}
\end{table}
\vspace{-0.5em}
We also measure decoding time-stamp accuracy of different models. The metric is the mean absolute error (MAE) between the start/end time-stamps of decoded hypothesized words and reference words, with incorrect words ignored. The reference time-stamps were obtained by aligning the audio with the reference using a bc-HMM system. In table \ref{tab:units_perf}, we report this metric as time-stamp-error (TSE, in ms) on \textit{dev-other}. We see that the ch-HMM model has the smallest TSE, confirming time-constraints in training targets helps the model to learn more accurate alignments.

\subsubsection{The effect of HMM topology}
\vspace{-0.5em}
Here we compare the impact of 1-state vs {\em chain} HMM topology for wp-HMM and ch-HMM models. For wp-HMM, in the {\em chain} case, the set of modeling units gets doubled from the 511 wordpieces as in the 1-state case. For ch-HMM, we choose a 3008-sized tree for the {\em chain} case, which is around two times of the 1632-leaves tree for the 1-state case. From Table \ref{chain}, we can see the impact on ch-HMM models is minor. However the impact on wp-HMM is obvious on \textit{dev-other}, where {\em chain} topology brings $5\%$ WER gain, which agrees with the finding in \cite{e2e_lfmmi}. We believe the reason behind the observation is that: The richer representation provided by {\em chain} topology, better modeling intra-class variations, contributes more to wordpiece units which are longer than chenones.

\begin{table}[!h]
    \centering
    \small
        \caption{\textit{dev-clean/other} MMI WER of wp-HMM and ch-HMM with 1-state and {\em chain} HMM topology} 
                \vspace{-0.2cm}
    \begin{tabular}{  p{1cm} | p{1.2cm} c | p{1.2cm} c}
 & \multicolumn{2}{c}{wp-HMM} & \multicolumn{2}{|c}{ch-HMM} \\
\cline{1-5}  
Topo. &  \small{1-state} &	 \small{{\em chain}}
&  \small{1-state} &	 \small{{\em chain}} \\
\hline
\hline
WER & 4.3 / 11.0 & 4.3 / 10.5 & 4.0 / 9.5 & 4.0 / 9.4 \\
\hline
\end{tabular}
\label{chain}
\end{table}
\vspace{-0.5em}
\subsubsection{The effect of denominator LM order}
\vspace{-0.5em}
Here we investigate the impact of denominator LM order on denominator FST size and training speed for wordpiece/chenone systems (wp-CTC/ch-HMM). From Table \ref{order} we can see that due to a large set of units which the den. LM is built upon, and the large CTC topology transducer (For a reference, den. FST w/ a 3-gram den. LM for wp-HMM is 5.2M), den. FST size in the wordpiece case is much larger than the chenone case, so that when increasing the order from 2 to 3, per-epoch training time increased by $110\%$, while it only increases by $12\%$ when changing order from 3 to 4 for ch-HMM. In terms of total training time, when increasing den. LM orders, wp-CTC training becomes much more expensive, while ch-HMM training even becomes cheaper. Considering the WER improvement for wp-CTC still looks worthwhile, we decide to stick with order 3 for wordpiece systems and 4 for chenone systems in other experiments. 

\begin{table}[]
    \centering
    \small
        \caption{\textit{dev-clean/other} MMI WER, denominator LM order/FST size, and training speed for wp-CTC and ch-HMM} 
                \vspace{-0.2cm}
    \begin{tabular}{  p{2cm} | c | c | c | c}
    & \multicolumn{2}{c|}{wp-CTC} & \multicolumn{2}{c}{ch-HMM} \\
\hline
den. LM order & 2 & 3 &  3 & 4 \\
den. FST size & 4.2MB & 10.2MB & 3.8MB & 4.6MB \\
\hline
\hline
WER &   4.8 / 11.4 &	4.4 / 10.6 & 4.3 / 9.9 &		4.0 / 9.5 \\
\hline
\# epochs & 112 & 121 & 84 & 54 \\
per-epoch hrs & 0.38 & 0.8 & 1 & 1.12 \\
\hline
\end{tabular}
\label{order}
\end{table}
\vspace{-0.5em}
\subsubsection{Benchmarking the 4 main systems with their optimal setup}
\vspace{-0.5em}
Here we conduct a comprehensive WER/RTF/TSE benchmark of the 4 main systems we have studied: wp-HMM, wp-CTC, ch-HMM, ch-CTC with their optimal training setup: optimal boost value for each, \textit{SpecAugment} \texttt{Large} policy for all, pre-training for all except wp-HMM, {\em chain} topology for wp/ch-HMM. From Table \ref{optimal}, we can see as expected, ch-HMM achieves the best TSE performance thanks to silence modeling and time-constraints used in training, ch-CTC achieves the best WER (thanks to blank+\textit{SpecAugment}), and also RTF. For wp-HMM and wp-CTC, they perform similarly well on RTF/WER (with wp-CTC's WER at stride 4 being a bit better), while wp-HMM's TSE is much better again thanks to silence modeling. This shows that wp-HMM, which doesn't rely on alignments, is an appealing choice when we need a large-stride $\&$ flat-start trained model providing accurate timestamps. Besides, though ch-CTC has worse TSE than ch-HMM (due to lack of time-constraints in training and CTC's peaky behavior), the gap is much smaller than that of wp-CTC/HMM, showing that silence modeling (which ch-CTC has but wp-CTC doesn't) can effectively improve time-stamp accuracy, even for CTC-based models. 

\begin{table}[!h]
    \centering
    \small
        \caption{\textit{dev-clean/other} bMMI WER/RTF/TSE of optimal systems}
    \begin{tabular}{  p{0.08\columnwidth} | c | c | c | c | c | c}
            \vspace{-0.2cm} 
            & \multicolumn{2}{c|}{wp-HMM} & \multicolumn{2}{c|}{wp-CTC} & ch-HMM & ch-CTC \\
\cline{1-7}
\hline
Stride & 8 & 4 & 8 & \multicolumn{3}{c}{4} \\
\hline
\hline
WER & 4.0/10.1 & 3.9/9.7 & 4.0/10.1 & 3.7/9.4 & 3.5/8.5 & 3.3/8.3 \\
\hline
RTF & 0.023 & 0.053 & 0.027 & 0.052 & 0.015 & 0.011 \\ 
\hline
TSE & 59 & 45 & 162 & 112 & 25 & 51 \\
\hline
\end{tabular}
\label{optimal}
\end{table}

\label{sec:pagestyle}
\vspace{-0.5em}
\subsection{CTC-bMMI training for real-world large-scale ASR tasks}
\vspace{-0.5em}
Here we apply the proposed CTC-bMMI training scheme with wordpiece/chenone units (i.e. wp-CTC-bMMI and ch-CTC-bMMI) in two real world large scale ASR tasks and compare with the corresponding ML baselines to confirm its effectiveness.  In the first application, we adopt wp-CTC-bMMI for training a large full-context Transformer model, for server-side ASR. In the second application, we adopt ch-CTC-bMMI for training a small limited-context streamable\footnote{Though the emphasis of our paper is bMMI for full-context ASR model training, we intentionally choose a limited-context scenario to show our method can work for streamable models as well.} Emformer \cite{emformer} using convolution operations similar to Conformer \cite{conformer}, for on-device ASR. We focus on CTC-bMMI rather than HMM-bMMI because the emphasis in the applications here is on WER rather than the token time-stamp accuracy.

\vspace{-0.5em}
\subsubsection{wp-CTC-bMMI for training large Transformer models}
\vspace{-0.5em}
\label{video_data}
Here, we compare CTC-bMMI with the standard CTC (i.e. CTC-ML) and RNN-T criteria on a real-world large scale English video ASR task. The training data consist of de-identified public videos with no personally identifiable information (PII), where only the audio part is used. Besides a development set, there are 3 test sets under different audio conditions: \textit{clean}, \textit{noisy} and \textit{extreme}. These test sets are further segmented by into audio chunks that are no longer than 45 seconds. Decoding is performed on these chunks unless otherwise specified. 
Training data are segmented into chunks with a maximum duration of 10s. Besides 39.4K hours of supervised training data (including two speed perturbed copies), we prepared 2.2M hours of unsupervised training data, with transcriptions obtained by decoding de-identified public videos by our internal ASR models. Several data filters are applied to keep the most useful data, e.g. confidence filter, word-per-second filter and country filter, etc. No human effort is involved in transcribing these unsupervised data. In total, we have 1.5M hours of semi-supervised training data. 

We use the same Transformer encoder architecture for each model, consisting of 24 layers, each with 12 attention heads, 768 embedding dimensions, and 3072 feed-forward dimensions. The encoder part has roughly 170M parameters. The input is the same as all other experiments: 80-dimensional log-Mel filter bank features at a 10ms frame rate. A stride of 8 is applied at the input layer by concatenating every 8 feature frames and then project to a dimension of 768, the same as the Transformer embedding dimension. For the RNN-T model, a predictor network consists of 512-dimensional embeddings for each token followed by two LSTM layers with 512 hidden nodes, then a linear projection to 1024-dimensional features before the joiner. For the joiner, the combined embeddings from the encoder and the predictor first go through a $tanh$ activation and then another linear projection to the target number of wordpieces. We use the same set of 511 wordpieces as modeling units for all models, and use the same $4$-gram LM for decoding CTC and CTC-bMMI models. In improve help convergence, for CTC(-ML) training we used CTC loss at intermediate layers. For RNN-T training we used CE loss at intermediate layers. The CTC-bMMI model is pre-trained by the CTC model. The boost value used is 2. Experiment results could be found in Table \ref{video}. We see that the CTC-bMMI model has large ($4-7\%$) WER improvements over CTC especially on the \textit{noisy} and \textit{extreme} sets and is almost on-par with the RNN-T model even without neural LM rescoring. RTF-wise, all models are similar. 

\begin{table}[h]
\vspace{-0.5em}
    \centering
    \small
        \caption{Comparing training criteria for Transformer-based ASR} 
        \vspace{-0.2cm}
    \begin{tabular}{ c || c  c  c  | c }
Loss & \textit{clean} & \textit{noisy} & \textit{extreme} & RTF \\
\cline{1-5}  
CTC  & 8.53 & 12.10 & 18.46 & 0.089 \\
CTC-bMMI  & 8.24 & 11.61 & 17.19 & 0.090 \\
RNN-T  & 8.01 & 11.49 & 17.04 & 0.094 \\
\hline
\end{tabular}
\vspace{-1em}

\label{video}
\end{table}
\vspace{-1em}
\subsubsection{ch-CTC-bMMI for training small Emformer models}
\vspace{-0.5em}
Here we study the effectiveness of CTC-bMMI using chenone modeling units in an on-device English ASR scenario, with CTC(-ML) as baselines. Training data are two subsets of the data used in Sec.  \ref{video_data}, containing $7000$ and $1000$ hours of videos correspondingly. We use the same test data as in Sec. \ref{video_data}. The model is an Emformer \cite{emformer} model supporting streaming speech recognition using block processing. In training, attention mask and ``right context hard copy'' are used to constrain the look ahead context for self-attention. In this experiment, each block consists of 1.4 seconds left context, 600 ms center chunk size, and 40 ms look-ahead context size. The algorithmic latency~\cite{emformer} of the acoustic model is 340 ms.  A stride of 4 is applied at the input layer by concatenating every 8 feature frames and then project to a dimension of 256, used as input to the stack of 12 Emformer layers. Each Emformer layer has a multi-head self-attention layer with four heads, input size 256, a feed-forward layer with hidden dim 1024, and a depth separable convolution layer with kernel size 15. The model has roughly 18M parameters. From the results in Table \ref{streaming}, we can see that in the 1000h condition, CTC-bMMI has $20-30\%$ relative WER improvement over CTC, which is much larger than the gain ($11-16\%$) in the 7000h condition, showing discriminative training helps more when we have less data, and when the models are smaller (comparing with Table \ref{video}).

\begin{table}[h]
    \centering
    \vspace{-0.5em}

    \small
        \caption{Comparing training criteria for Emformer-based ASR} 
        \vspace{-0.2cm}
    \begin{tabular}{ c || c  c  c  | c }
Criterion & \textit{clean} & \textit{noisy} & \textit{extreme} & training hours \\
\cline{1-5}  
CTC  & 25.38 & 32.18 & 39.65 & \multirow{2}{*}{1000h}  \\
CTC-bMMI  & 17.63 & 23.36 & 31.02 &  \\
\hline
CTC  & 18.44 & 23.97 & 31.38 &  \multirow{2}{*}{7000h} \\
CTC-bMMI  & 15.45& 20.59 & 27.71 &  \\
\hline
\end{tabular}
\label{streaming}
\end{table}
\vspace{-1em}
\subsection{bc-HMM-MMI for alignment model training}
\vspace{-0.5em}

Here, we study an important application of HMM-MMI models with bi-char units (bc-HMM-MMI): alignment generation. Accurate alignments are important for ASR, in terms of both audio segmentation and providing training targets for main/auxiliary ASR training tasks even for RNN-T \cite{chunxi2021aux,ar_rnnt}. In order to train an alignment model from scratch, people have been mainly relying on GMM-HMMs, e.g. from Kaldi \cite{kaldi}. However, single-stage trained, HMM-based neural models, e.g. bc-HMM-MMI models, can be more appealing candidates (used in Kaldi OCR recipes\cite{arora2019using} already), which may provide more accurate alignments especially on noisy data, and moreover, enable an all-neural acoustic modeling pipeline. To the best of our knowledge, there's no prior literature confirming this by benchmarking bc-HMM-MMI models with GMM-HMMs. Here we conduct this benchmark by training a bc-HMM-MMI neural model and a GMM model (following the Kaldi recipe) with the same data and graphemic lexicon, evaluate their WERs, and then generate alignments on the same training data, on top which we then train two CE neural models and evaluate their WERs for measuring the alignment quality. The two CE models and the bc-HMM-MMI alignment model all have the same architecture as the one used in \ref{libri_setup}, except the stride is 3 here. Using the same architecture enables us to show another advantage of bc-HMM-MMI alignment models: Besides generating alignments, it can also serve as a pre-trained seed model for the following modeling stage to improve training performance, which can't be done with GMMs. We conduct the experiments on Librispeech where we train models on the full 960h data and evaluate WERs on \textit{dev-other}, and a Tagalog Video ASR task (whose description is the same as \ref{video_data}) where we train models on 1000h Tagalog videos and evaluate WERs on the \textit{noisy} test set. From the results shown in Table \ref{bootstrap}, we can see that the bc-HMM-MMI neural alignment model achieves on-par alignment quality as GMM evaluated by CE WER. On Tagalog Video ASR, which is much noiser than Librispeech, the bc-HMM-MMI model is capable of generating much better alignmetns, reducing CE WER by $14\%$ relatively. Besides, pre-traing the CE models with bc-HMM-MMI seed models indeed bring down CE WERs further, by $2\%$ (Tagalog) or $5\%$ (Librispeech) relatively. This shows besides serving as a strong alignment model, a bc-HMM-MMI model can also serve as as a seed model for downstream modeling tasks.

\begin{table}[]
    \centering
    \small
        \caption{Alignment model and CE model WERs, on Tagalog video (\textit{noisy}) and Librispeech (\textit{dev-other})}
    \begin{tabular}{  c | c |  c | c  | c}
 & \multicolumn{2}{c|}{Alignment Model} & CE & CE w/ seed \\
\hline
\hline
 \multirow{2}{*}{Tagalog Video}  & GMM & 61.7 &	38.0 & -\\
 & bc-HMM-MMI  & 27.5 & 32.6 & 31.9 \\
\hline
\multirow{2}{*}{Librispeech} & GMM & 30.1 & 11.3 & - \\
& bc-HMM-MMI  & 10.0 & 11.2 & 10.6 \\
\hline
\end{tabular}

\label{bootstrap}
\end{table}

\vspace{-0.5em}
\section{Conclusion}
\vspace{-0.5em}
In this paper, we generalized the original chunk-wise HMM-based LF-bMMI training framework to a new framework, where full-context neural network training is enabled by full-sequence LF-bMMI training, supporting both HMM and CTC as the label topology, and mono-char/bi-char/chenone/wordpieces as modeling units. Comprehensive studies were conducted on Librispeech to understand the impact of boost factor, CE/ML pre-training, \textit{SpecAugment} and denominator LM order to different training schemes. From this framework, we proposed wp-CTC-bMMI and ch-CTC-bMMI training schemes with WER advantages, studied also in two large scale real-world ASR tasks, and wp-HMM-bMMI training scheme with advantages in large-stride inference, time-stamps accuracy, and alignment-free training. In the future we would like to further generalize LF-bMMI training to RNN-T-type of topologies.


\bibliographystyle{IEEEbib}
\bibliography{refs}

\begin{thebibliography}{10}

\bibitem{hmm_dnn}
George~E. Dahl, Dong Yu, Li~Deng, and Alex Acero,
\newblock ``Context-dependent pre-trained deep neural networks for
  large-vocabulary speech recognition,''
\newblock {\em IEEE Transactions on Audio, Speech, and Language Processing},
  vol. 20, no. 1, pp. 30--42, 2012.

\bibitem{bahl_mmie}
L.~Bahl, P.~Brown, P.~de~Souza, and R.~Mercer,
\newblock ``Maximum mutual information estimation of hidden markov model
  parameters for speech recognition,''
\newblock in {\em ICASSP '86. IEEE International Conference on Acoustics,
  Speech, and Signal Processing}, 1986, vol.~11, pp. 49--52.

\bibitem{Vesely_13}
K.~{Vesely}, M.~{Hannemann}, and L.~{Burget},
\newblock ``Semi-supervised training of deep neural networks,''
\newblock in {\em ASRU 2013}.

\bibitem{saon2012discriminative}
George Saon and Brian Kingsbury,
\newblock ``Discriminative feature-space transforms using deep neural
  networks,''
\newblock in {\em Thirteenth Annual Conference of the International Speech
  Communication Association}, 2012.

\bibitem{lfmmi}
Daniel Povey, Vijayaditya Peddinti, Daniel Galvez, Pegah Ghahremani, Vimal
  Manohar, Xingyu Na, Yiming Wang, and Sanjeev Khudanpur,
\newblock ``Purely sequence-trained neural networks for asr based on
  lattice-free mmi,''
\newblock in {\em Interspeech}, 2016.

\bibitem{e2e_lfmmi}
Hossein Hadian, Hossein Sameti, Daniel Povey, and Sanjeev Khudanpur,
\newblock ``Flat-start single-stage discriminatively trained hmm-based models
  for asr,''
\newblock {\em IEEE/ACM Transactions on Audio, Speech, and Language
  Processing}, vol. 26, no. 11, pp. 1949--1961, 2018.

\bibitem{ctc}
Alex Graves, Santiago Fernández, and Faustino Gomez,
\newblock ``Connectionist temporal classification: Labelling unsegmented
  sequence data with recurrent neural networks,''
\newblock in {\em ICML 2006}.

\bibitem{graves2012rnnt}
Alex Graves,
\newblock ``Sequence transduction with recurrent neural networks,''
\newblock {\em arXiv preprint arXiv:1211.3711}, 2012.

\bibitem{bahdanau2016attention}
Dzmitry Bahdanau, Jan Chorowski, Dmitriy Serdyuk, Philémon Brakel, and Yoshua
  Bengio,
\newblock ``End-to-end attention-based large vocabulary speech recognition,''
\newblock in {\em 2016 IEEE International Conference on Acoustics, Speech and
  Signal Processing (ICASSP)}, 2016, pp. 4945--4949.

\bibitem{las}
W.~{Chan}, N.~{Jaitly}, Q.~{Le}, and O.~{Vinyals},
\newblock ``Listen, attend and spell: A neural network for large vocabulary
  conversational speech recognition,''
\newblock in {\em ICASSP}, 2016.

\bibitem{chiu2018seq2seq}
Chung-Cheng Chiu, Tara~N. Sainath, Yonghui Wu, Rohit Prabhavalkar, Patrick
  Nguyen, Zhifeng Chen, Anjuli Kannan, Ron~J. Weiss, Kanishka Rao, Ekaterina
  Gonina, Navdeep Jaitly, Bo~Li, Jan Chorowski, and Michiel Bacchiani,
\newblock ``State-of-the-art speech recognition with sequence-to-sequence
  models,''
\newblock in {\em 2018 IEEE International Conference on Acoustics, Speech and
  Signal Processing (ICASSP)}, 2018, pp. 4774--4778.

\bibitem{xiaohui2021benchmarking}
Xiaohui Zhang, Frank Zhang, Chunxi Liu, Kjell Schubert, Julian Chan, Pradyot
  Prakash, Jun Liu, Ching-Feng Yeh, Fuchun Peng, Yatharth Saraf, and Geoffrey
  Zweig,
\newblock ``Benchmarking lf-mmi, ctc and rnn-t criteria for streaming asr,''
\newblock in {\em 2021 IEEE Spoken Language Technology Workshop (SLT)}, 2021,
  pp. 46--51.

\bibitem{graves2013rnn}
Alex Graves, Abdel-rahman Mohamed, and Geoffrey Hinton,
\newblock ``Speech recognition with deep recurrent neural networks,''
\newblock in {\em 2013 IEEE International Conference on Acoustics, Speech and
  Signal Processing}, 2013, pp. 6645--6649.

\bibitem{chunxi2021aux}
Chunxi Liu, Frank Zhang, Duc Le, Suyoun Kim, Yatharth Saraf, and Geoffrey
  Zweig,
\newblock ``Improving rnn transducer based asr with auxiliary tasks,''
\newblock in {\em 2021 IEEE Spoken Language Technology Workshop (SLT)}, 2021,
  pp. 172--179.

\bibitem{ar_rnnt}
Jay Mahadeokar, Yuan Shangguan, Duc Le, Gil Keren, Hang Su, Thong Le,
  Ching-Feng Yeh, Christian Fuegen, and Michael~L Seltzer,
\newblock ``Alignment restricted streaming recurrent neural network
  transducer,''
\newblock in {\em 2021 IEEE Spoken Language Technology Workshop (SLT)}. IEEE,
  2021, pp. 52--59.

\bibitem{Zeyer2020transducer}
Albert Zeyer, André Merboldt, Ralf Schlüter, and Hermann Ney,
\newblock ``{A New Training Pipeline for an Improved Neural Transducer},''
\newblock in {\em Proc. Interspeech 2020}, 2020, pp. 2812--2816.

\bibitem{semisup_lfmmi}
V.~{Manohar}, H.~{Hadian}, D.~{Povey}, and S.~{Khudanpur},
\newblock ``Semi-supervised training of acoustic models using {L}attice-{F}ree
  {MMI},''
\newblock in {\em ICASSP}, 2018.

\bibitem{ctc_crf}
Hongyu Xiang and Zhijian Ou,
\newblock ``Crf-based single-stage acoustic modeling with ctc topology,''
\newblock in {\em ICASSP 2019 - 2019 IEEE International Conference on
  Acoustics, Speech and Signal Processing (ICASSP)}, 2019, pp. 5676--5680.

\bibitem{duc2021rnnt_context}
Duc Le, Mahaveer Jain, Gil Keren, Suyoun Kim, Yangyang Shi, Jay Mahadeokar,
  Julian Chan, Yuan Shangguan, Christian Fuegen, Ozlem Kalinli, Yatharth Saraf,
  and Michael~L. Seltzer,
\newblock ``Contextualized streaming end-to-end speech recognition with
  trie-based deep biasing and shallow fusion,''
\newblock {\em CoRR}, vol. abs/2104.02194, 2021.

\bibitem{le2019senones}
Duc Le, Xiaohui Zhang, Weiyi Zheng, Christian F{\"u}gen, Geoffrey Zweig, and
  Michael~L Seltzer,
\newblock ``From senones to chenones: Tied context-dependent graphemes for
  hybrid speech recognition,''
\newblock {\em ASRU}, 2019.

\bibitem{ctc_smbr}
Haşim Sak, Andrew Senior, Kanishka Rao, Ozan İrsoy, Alex Graves, Françoise
  Beaufays, and Johan Schalkwyk,
\newblock ``Learning acoustic frame labeling for speech recognition with
  recurrent neural networks,''
\newblock in {\em 2015 IEEE International Conference on Acoustics, Speech and
  Signal Processing (ICASSP)}, 2015, pp. 4280--4284.

\bibitem{povey2008bmmi}
Daniel Povey, Dimitri Kanevsky, Brian Kingsbury, Bhuvana Ramabhadran, George
  Saon, and Karthik Visweswariah,
\newblock ``Boosted mmi for model and feature-space discriminative training,''
\newblock in {\em 2008 IEEE International Conference on Acoustics, Speech and
  Signal Processing}, 2008, pp. 4057--4060.

\bibitem{chen2018sequence}
Zhehuai Chen, Yanmin Qian, and Kai Yu,
\newblock ``Sequence discriminative training for deep learning based acoustic
  keyword spotting,''
\newblock {\em Speech Communication}, vol. 102, pp. 100--111, 2018.

\bibitem{weng2019lfbmmi}
Chao Weng and Dong Yu,
\newblock ``A comparison of lattice-free discriminative training criteria for
  purely sequence-trained neural network acoustic models,''
\newblock in {\em ICASSP 2019-2019 IEEE International Conference on Acoustics,
  Speech and Signal Processing (ICASSP)}. IEEE, 2019, pp. 6430--6434.

\bibitem{baskar2019promising}
Murali~Karthick Baskar, Luk{\'a}{\v{s}} Burget, Shinji Watanabe, Martin
  Karafi{\'a}t, Takaaki Hori, and Jan~Honza {\v{C}}ernock{\`y},
\newblock ``Promising accurate prefix boosting for sequence-to-sequence asr,''
\newblock in {\em ICASSP 2019-2019 IEEE International Conference on Acoustics,
  Speech and Signal Processing (ICASSP)}. IEEE, 2019, pp. 5646--5650.

\bibitem{smithsoftmax}
Kevin Gimpel Noah~A Smith,
\newblock ``Softmax-margin training for structured log-linear models,''
\newblock .

\bibitem{chen2015pronunciation}
Guoguo Chen, Hainan Xu, Minhua Wu, Daniel Povey, and Sanjeev Khudanpur,
\newblock ``Pronunciation and silence probability modeling for asr,''
\newblock in {\em Sixteenth Annual Conference of the International Speech
  Communication Association}, 2015.

\bibitem{zeyer2021does}
Albert Zeyer, Ralf Schl{\"u}ter, and Hermann Ney,
\newblock ``Why does ctc result in peaky behavior?,''
\newblock {\em arXiv preprint arXiv:2105.14849}, 2021.

\bibitem{young_tied_states}
S.~J. Young, J.~J. Odell, and P.~C. Woodland,
\newblock ``Tree-based state tying for high accuracy acoustic modelling,''
\newblock in {\em Proceedings of the Workshop on Human Language Technology},
  USA, 1994, HLT '94, p. 307–312, Association for Computational Linguistics.

\bibitem{spm}
Taku Kudo and John Richardson,
\newblock ``{S}entence{P}iece: A simple and language independent subword
  tokenizer and detokenizer for neural text processing,''
\newblock in {\em Proceedings of the 2018 Conference on Empirical Methods in
  Natural Language Processing: System Demonstrations}, Brussels, Belgium, Nov.
  2018, pp. 66--71, Association for Computational Linguistics.

\bibitem{kudo2018sentencepiece}
Taku Kudo and John Richardson,
\newblock ``Sentencepiece: A simple and language independent subword tokenizer
  and detokenizer for neural text processing,''
\newblock {\em arXiv preprint arXiv:1808.06226}, 2018.

\bibitem{zhang2020wp_hybrid}
Frank Zhang, Yongqiang Wang, Xiaohui Zhang, Chunxi Liu, Yatharth Saraf, and
  Geoffrey Zweig,
\newblock ``{Faster, Simpler and More Accurate Hybrid ASR Systems Using
  Wordpieces},''
\newblock in {\em Proc. Interspeech 2020}, 2020, pp. 976--980.

\bibitem{manohar2015semi}
Vimal Manohar, Daniel Povey, and Sanjeev Khudanpur,
\newblock ``Semi-supervised maximum mutual information training of deep neural
  network acoustic models,''
\newblock in {\em Sixteenth Annual Conference of the International Speech
  Communication Association}, 2015.

\bibitem{librispeech}
V.~{Panayotov}, G.~{Chen}, D.~{Povey}, and S.~{Khudanpur},
\newblock ``Librispeech: An asr corpus based on public domain audio books,''
\newblock in {\em ICASSP}, 2015.

\bibitem{lstm}
Sepp Hochreiter and J{\"u}rgen Schmidhuber,
\newblock ``Long short-term memory,''
\newblock {\em Neural computation}, vol. 9, no. 8, pp. 1735--1780, 1997.

\bibitem{tdnn}
Kevin~J Lang, Alex~H Waibel, and Geoffrey~E Hinton,
\newblock ``A time-delay neural network architecture for isolated word
  recognition,''
\newblock {\em Neural networks}, vol. 3, no. 1, pp. 23--43, 1990.

\bibitem{vijay_tdnn}
Vijayaditya Peddinti, Daniel Povey, and Sanjeev Khudanpur,
\newblock ``A time delay neural network architecture for efficient modeling of
  long temporal contexts,''
\newblock in {\em Sixteenth Annual Conference of the International Speech
  Communication Association}, 2015.

\bibitem{spec_augment}
Daniel~S Park, William Chan, Yu~Zhang, Chung-Cheng Chiu, Barret Zoph, Ekin~D
  Cubuk, and Quoc~V Le,
\newblock ``Specaugment: A simple data augmentation method for automatic speech
  recognition,''
\newblock {\em arXiv preprint arXiv:1904.08779}, 2019.

\bibitem{emformer}
Yangyang Shi, Yongqiang Wang, Chunyang Wu, Ching-Feng Yeh, and Others,
\newblock ``{Emformer: Efficient Memory Transformer Based Acoustic Model For
  Low Latency Streaming Speech Recognition},''
\newblock in {\em Proc. ICASSP}, 2021.

\bibitem{conformer}
Anmol Gulati, James Qin, Chung~Cheng Chiu, and Others,
\newblock ``{Conformer: Convolution-augmented transformer for speech
  recognition},''
\newblock in {\em Proc. INTERSPEECH}, 2020.

\bibitem{kaldi}
Daniel Povey, Arnab Ghoshal, Gilles Boulianne, Lukas Burget, Ondrej Glembek,
  Nagendra Goel, Mirko Hannemann, Petr Motlicek, Yanmin Qian, Petr Schwarz, Jan
  Silovsky, Georg Stemmer, and Karel Vesely,
\newblock ``The kaldi speech recognition toolkit,''
\newblock in {\em IEEE 2011 Workshop on Automatic Speech Recognition and
  Understanding}. Dec. 2011, IEEE Signal Processing Society,
\newblock IEEE Catalog No.: CFP11SRW-USB.

\bibitem{arora2019using}
Ashish Arora, Chun~Chieh Chang, Babak Rekabdar, Bagher BabaAli, Daniel Povey,
  David Etter, Desh Raj, Hossein Hadian, Jan Trmal, Paola Garcia, et~al.,
\newblock ``Using asr methods for ocr,''
\newblock in {\em 2019 International Conference on Document Analysis and
  Recognition (ICDAR)}. IEEE, 2019, pp. 663--668.

\end{thebibliography}

\end{document}